\documentclass[review]{elsarticle}
\usepackage{lineno,hyperref}
\usepackage{comment}
\modulolinenumbers[5]
\journal{Journal of \LaTeX\ Templates}

\begin{document}

\begin{frontmatter}

\title{Non-mesonic decay of the $\eta$-mesic $^{3}\hspace{-0.03cm}\mbox{He}$ via
 $pd\rightarrow(^{3}\hspace{-0.03cm}\mbox{He}$-$\eta)_{bound}\rightarrow$ $^{3}\hspace{-0.03cm}\mbox{He} 2\gamma (6\gamma$) reaction}

\author[uj,infn]{M. Skurzok}
\ead{magdalena.skurzok@uj.edu.pl}

\author[nara]{S. Hirenzaki}
\author[nara]{S. Kinutani}
\author[nara]{H. Konishi}
\author[uj]{P. Moskal}
\author[nara,osaka]{H. Nagahiro}

\author[uj]{O. Rundel}

\address[uj]{Institute of Physics, Jagiellonian University, prof. Stanis{\l}awa {\L}ojasiewicza str.~11, 30-348 Krak\'{o}w, Poland}
\address[infn]{INFN, Laboratori Nazionali di Frascati, 00044 Frascati, Italy}
\address[nara]{Department of Physics, Nara Women's University, Nara 630-8506, Japan}
\address[osaka]{Research Center for Nuclear Physics (RCNP), Osaka University, Ibaraki 567-0047, Japan}

\begin{abstract}
In this article a theoretical model for the $\eta$-mesic $^{3}\hspace{-0.03cm}\mbox{He}$ non-mesonic decay channels is presented. We present the resultant relative momentum distribution of bound $^{3}\hspace{-0.03cm}\mbox{He}$-$\eta$ as well as in-medium branching ratios of $\eta\rightarrow 2\gamma$ and $\eta\rightarrow 3\pi^0$, which are crucial for the Monte Carlo simulations of measured processes and thus for the experimental data interpretation. As an example we also apply the model for the estimation of the detection efficiency of the WASA-at-COSY detector.
\end{abstract}


\end{frontmatter}


\section{Introduction}



In the surrounding universe, aside from usual matter like atomic nuclei, a variety of uncommon exotic objects can be found. Although many of them, such as hypernuclei~\cite{Danysz}, tetraquarks~\cite{Tetraq}, pentaquarks~\cite{Pentaquark} or dibaryons~\cite{dibar_jeden,dibar_dwaa,dibar_trzy} have already been discovered and investigated, still many exotic systems are theoretically predicted but never experimentally confirmed. 
Mesic nuclei, consisting of a nucleus bound via the strong interaction with a neutral meson are one example. They are currently one of the hottest topics in nuclear and hadronic physics, both from experimental~\cite{Magda_PLB2018,Skurzok_NPA,Adlarson_2013,Tanaka,Machner_2015,Metag2017} and theoretical~\cite{BassMos,Ikeno_EPJ2017,Xie2019,Xie2017,Fix2017,Barnea2017,Gal2017,Gal2015,Friedman,Kelkar_2016_new,Kelkar,Kelkar_new,Wilkin_2016,Wilkin2,BassTom2006,BassTom,Hirenzaki1,Nagahiro_2008rj,Nagahiro_2013,Hirenzaki_2010,WycechKrzemien,Niskanen_2015} standpoints. Some of the most promising candidates for such bound states are $\eta(\eta')$-mesic nuclei since the $\eta$-nucleon interaction was found to be attractive~\cite{BhaleraoLiu,HaiderLiu1} while the imaginary $\eta'$-nucleus potential for near threshold is significantly smaller than the modulus of the real part~\cite{Nanova2018}.

~Recently performed studies of hadron- and photo-production of the $\eta$ meson result in a wide range of values of the $\eta N$ scattering length indicating that the $\eta$ meson-nucleon interaction is attractive and strong enough to create even light $\eta$-nucleus bound systems~\cite{Xie2019,Xie2017,Fix2017,Barnea2017,Gal2017,Green,Wilkin1,WycechGreen}. However, none of the experiments performed till now have found a clear signature confirming their 
existence. They only provided signals which might be interpreted as indications of the 
$\eta$-mesic nuclei as well as allow the determination of the upper limits of the total cross section for the bound state formation~\cite{Magda_PLB2018,Skurzok_NPA,Adlarson_2013,Machner_2015,Metag2017,Kelkar,Wilkin_2016,Berger,Mayer,Sokol_2001,Smyrski1,Mersmann,Budzanowski,Papenbrock,PMActa,HaiderLiu,Krusche_Wilkin,Moskal_FewBody,Acta_2016,Moskal_Acta2016,Moskal_AIP2017}. 

One of the most recent and promising experiments related to $\eta$-mesic Helium nuclei have been performed by the WASA-at-COSY Collaboration~\cite{Magda_PLB2018,Skurzok_NPA,Adlarson_2013,Wilkin_epj2017}. The measurements were carried out with high statistics with the WASA detection setup in deuteron-deuteron ($^{4}\hspace{-0.03cm}\mbox{He}$-$\eta$)~\cite{Magda_PLB2018,Skurzok_NPA,Adlarson_2013} and proton-deuteron ($^{3}\hspace{-0.03cm}\mbox{He}$-$\eta$)~\cite{Skurzok_epj2018,Rundel_lum2017} fusion reactions using the ramped beam technique. 

The analysis dedicated to search for $^{4}\hspace{-0.03cm}\mbox{He}$-$\eta$ mesic nuclei in $dd\rightarrow$ $^{3}\hspace{-0.03cm}\mbox{He} n \pi{}^{0}$ and $dd\rightarrow$ $^{3}\hspace{-0.03cm}\mbox{He} p \pi^{-}$ processes resulted in the upper limits of the total cross section at a 90\% confidence level equal to roughly 3~nb and 6~nb, respectively~\cite{Skurzok_NPA}. The determined excitation functions were compared with the predictions of the model proposed in Ref.~\cite{Ikeno_EPJ2017}, this allowed to put a constraint on the $\eta$-$^4$He optical potential parameters~\cite{Magda_PLB2018}.

The analyses and interpretations of all experiments up till now have been performed assuming a mechanism according to which (after the $\eta$-mesic nucleus creation) the $\eta$ meson is absorbed on one of the nucleons inside helium and may propagate in the nucleus via consecutive excitations of nucleons to the $N^{*}$(1535) state, until the resonance decays into the nucleon-pion pair. Thus far, Monte Carlo simulations, used for the estimation of the detector systems registration efficiency, have been carried out assuming that the $N^{*}$ momentum distribution is the same as the distribution of nucleons~\cite{Nogga}. Only recently the first model describing the $N^{*}$ momentum in the $N^{*}$-$^{3}\hspace{-0.03cm}\mbox{He}$ bound state was proposed in references~\cite{Kelkar_2016_new,Kelkar_2015_new}. Another theoretical model predicting non-$N \pi$ decays of the mesic-helium formed via two-nucleon absorption process was pointed in Refs~\cite{WycechKrzemien,Wilkin_2014}.

Recently, a new mechanism of the hypothetical $\eta$-mesic helium decay was considered, namely via $\eta$ meson decay while it is still "orbiting" around a nucleus. In order to avoid complications due to final state interactions neutral decay channels such as $\eta\to 2\gamma$ and $\eta \to 3\pi^0 \to 6\gamma$ constituting more than 70\% of $\eta$ meson decays~\cite{Pdg2018} are best suited for these studies. The dedicated measurement was performed for the first time with the WASA-at-COSY facility to search for $\eta$-mesic $^{3}\hspace{-0.03cm}\mbox{He}$ in $pd \rightarrow$ $^{3}\hspace{-0.03cm}\mbox{He} 2\gamma$ and $pd \rightarrow$ $^{3}\hspace{-0.03cm}\mbox{He} 6\gamma$ reactions. 

In this article we present a theoretical model for the $\eta$-mesic $^{3}\hspace{-0.03cm}\mbox{He}$ non-mesonic decay channels. We present the resulting relative momentum distribution of bound $^{3}\hspace{-0.03cm}\mbox{He}$-$\eta$ as well as in-medium branching ratios of $\eta\rightarrow 2\gamma$ and $\eta\rightarrow 3\pi^0$,  which are crucial for the Monte Carlo simulations of measured processes and hence for the experimental data interpretation. As an example we apply the model for the estimation of the detection efficiency of the WASA-at-COSY detector.  

Our goal is to know the $\eta$-nucleus interaction and hence the (non)existence of the quasi-stable bound states. The criterion of the existence of quasi-stable bound states with long enough lifetime can be verified by the measurement of binding energy and the width of the state and by experimental determination of the in-medium branching ratios of $\eta$ decay into $2\gamma$ and $3\pi^0$.

\section{Theoretical model}

\noindent 

\begin{figure}[h!]
\centering
\includegraphics[width=14.0cm,height=6.0cm]{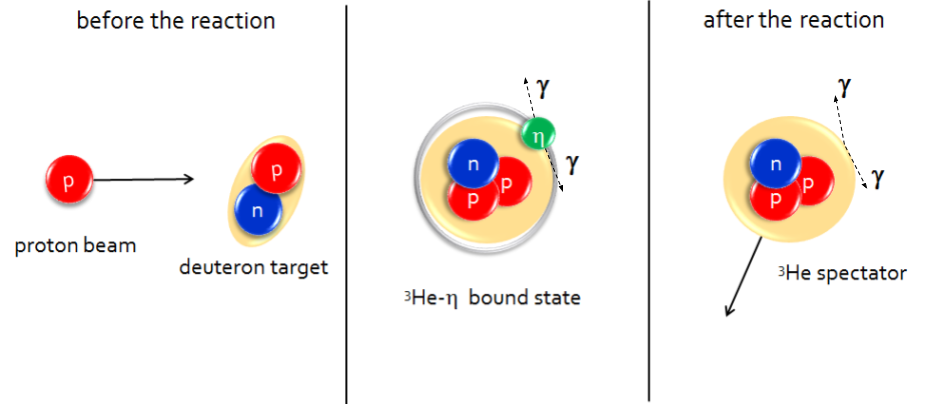}
\caption{Scheme of the $^{3}\hspace{-0.03cm}\mbox{He}$-$\eta$ mesic nucleus production and decay in $pd \rightarrow$ $^{3}\hspace{-0.03cm}\mbox{He} 2\gamma$ reaction.\label{free_reaction}}
\end{figure}

A theoretical model has been developed in order to describe the kinematics of the decay of $\eta$-mesic $^{3}\hspace{-0.03cm}\mbox{He}$
nucleus in $pd \rightarrow$ $^{3}\hspace{-0.03cm}\mbox{He} 2\gamma$ and
$pd \rightarrow$ $^{3}\hspace{-0.03cm}\mbox{He} 6\gamma$
processes. The proposed mechanism proceeds according to the scheme
shown in Fig.~\ref{free_reaction}.  
\noindent
The proton-deuteron collision leads to the creation of
$^{3}\hspace{-0.03cm}\mbox{He}$ nucleus bound with the $\eta$ meson via
strong interaction. The bound $\eta$ meson is orbiting inside the nucleus
until it decays into $2\gamma$ or $3\pi^{0}$, wheres $3\pi^{0}$ decay almost immediately to $6\gamma$. The $^{3}\hspace{-0.03cm}\mbox{He}$
nucleus plays the role of a spectator in the decay
processes. It is assumed, that the $\eta$ meson is moving with Fermi momentum inside mesic nucleus before decaying into $2\gamma$ or $3\pi^0$.
The Fermi momentum distribution is evaluated based on the calculated $\eta$ wavefunction
described in detail in 
the next subsection. 

\subsection{Structure of the $\eta$ bound state and in-medium branching ratios of $\eta\rightarrow 2\gamma$ and $\eta\rightarrow 3\pi^0$}

The structure of the hypothetical $\eta$ bound state produced in $pd \rightarrow$
$^{3}\hspace{-0.03cm}\mbox{He} \eta$ reaction can be
described as the solution of the Klein-Gordon equation: 

\begin{equation}~\label{KG_eq}
\left[-\vec{\bigtriangledown}^{2}+\mu^{2}+2\mu
 U_{\rm opt}(r)\right]\psi(\vec{r})=E^{2}_{\rm KG}\psi(\vec{r}), 
\label{eq_1}
\end{equation}
with $E_{\rm KG}$ and $\mu$ denoting the Klein-Gordon energy and $^3$He-$\eta$ reduced mass, respectively. The $U_{\rm opt}(r)$ is the optical potential describing the interaction between $^3$He and $\eta$, and is assumed to have the functional form:
%
%
%
%
%
\begin{equation}
U_{\rm opt}(r)=(V_{0}+iW_{0})\frac{\rho(r)}{\rho_0} \ ,
\end{equation}
where $\rho(r)$ is the density distribution of $^3$He and $\rho_0$ is
the normal nuclear density $\rho_0=0.17$~fm$^{-3}$.  The $^3$He density
distribution $\rho(r)$ is obtained by the theoretical calculation
described in Refs.~\cite{Hiyama1,Hiyama2,Hiyama3}. 
%
%

The equation is solved numerically for several sets of
real $(V_0)$ and imaginary $(W_0)$ optical potential parameters
to obtain the Klein-Gordon energy $E_{\rm KG}$ and the wavefunction $\psi(\vec{r})$. The binding energy $B_{\rm s}$ and the nuclear absorption width
$\Gamma_{\rm abs}$ of the $\eta$ bound state are defined via the
Klein-Gordon energy $E_{\rm KG}$ as $B_{\rm s}={\rm Re}(\mu-E_{\rm KG})$ and 
$\Gamma_{\rm abs}=-2{\rm Im}(E_{\rm KG})$, respectively.  
We show the assumed strength of the optical potential parameters
($V_0$ and $W_0$), the obtained binding energy $B_{\rm s}$, and
nuclear absorption width $\Gamma_{\rm abs}$ in Table~\ref{table_1}.  In
addition to the strongly absorptive potential with $W_0=-20$~MeV, we
also assumed the weakly absorptive potential with $W_0=-1$~MeV as
indicated in Ref.~\cite{Magda_PLB2018} for the
$^4$He-$\eta$ system.\\

\begin{table}[h!]
\begin{normalsize}
\begin{center}
\begin{tabular}{|c|c|c|c|}\hline
$(V_{0},W_{0})$ [MeV]  &($B_{\rm s}$, $\Gamma_{\rm abs}$) [MeV] &
 BR$^*_{\eta\rightarrow 2\gamma}$ & BR$^*_{\eta\rightarrow 3\pi^0}$\\
\hline 
$-$(75,20) &(4.06, 15.66) & $3.30 \times 10^{-5}$ & $2.73\times
	     10^{-5}$   \\
 $-$(90,20) &(11.16, 20.65)& $2.50 \times 10^{-5}$ & $2.07\times 10^{-5} $
  \\
$-$(75,1) &(5.96, 0.76) & $6.78\times 10^{-4}$&$5.62\times 10^{-4}$\\
$-$(90,1) &(12.67, 1.02)& $5.06\times 10^{-4}$& $4.20\times 10^{-4}$\\
\hline
\end{tabular}
\end{center}
\begin{center}
 \caption{The binding energies $B_{\rm s}$ and nuclear absorption widths
 $\Gamma_{\rm abs}$ values for the $^3$He-$\eta$ ground $(0s)$ states
 obtained by solving Eq.~(\ref{eq_1}) are listed with the optical
 potential parameters $(V_0,W_0)$ assumed in the present calculation.
 Evaluated in-medium branching ratios BR$^*$ are also shown  (See
 details in text).
 \label{table_1}} 
\end{center}
\end{normalsize}
\end{table}

\vspace{-0.8cm}

\begin{figure}[h!]
\centering
  \includegraphics[width=10.0cm,height=7.0cm]{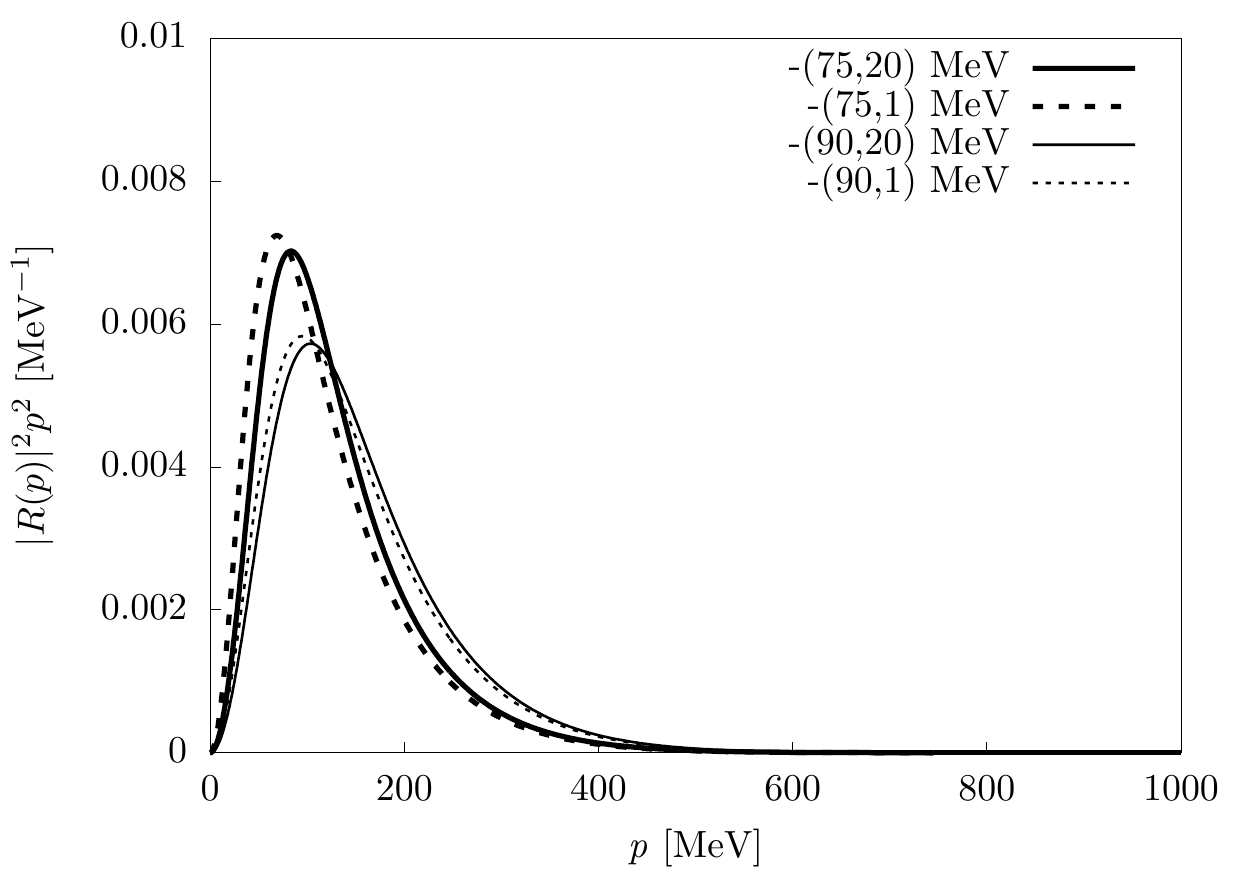}
\vspace{-0.5cm}
\caption{Fermi momentum distribution of the $\eta$ meson in
 $^{3}\hspace{-0.03cm}\mbox{He}$-$\eta$ bound system estimated for
 ($V_{0},W_{0}$)=$-$(75,20)~MeV (thick solid line),
 ($V_{0},W_{0}$)=$-$(75,1)~MeV (thick dotted line),
 ($V_{0},W_{0}$)=$-$(90,20)~MeV (thin solid line), and
 ($V_{0},W_{0}$)=$-$(90,1)~MeV (thin dotted line).
 The distributions are
 normalized to be $1$ in the whole momentum range.\label{fig_Fermi}
}
\end{figure}

By transforming the coordinate space wavefunction $\psi(\vec{r})$
obtained by solving Eq.~(\ref{eq_1}), we can derive the momentum space wavefunction in the form $\phi(\vec{p})=R(p)Y_{\ell m}(\hat{p})$ using:
\begin{equation}
 \phi(\vec{p})=\frac{1}{(2\pi)^{3/2}}\int
  e^{i\vec{p}\cdot\vec{r}}\psi(\vec{r}) d\vec{r} \ ,
  \label{eq:mom_wave}
\end{equation}
and we can evaluate the relative $^3$He-$\eta$ momentum distribution
using $|R(p)|^2$, 
where $R(p)$ satisfies the normalization condition $\displaystyle
\int |R(p)|^2p^2 dp=1$. 
%
%
%
%
%
The momentum distributions $|R(p)|^2p^2$ for four sets of potential
parameter values are presented in Fig.~\ref{fig_Fermi}.
  We also derive the momentum space wavefunction of the nucleon
  $\phi_N(\vec{p})=R_N(p)Y_{00}(\hat{p})$ in the same manner using the
  coordinate space $s$-wave nucleon wavefunction $\Psi_N(\vec{r})$,
  which is assumed to be related to the nucleon density distribution
  $\rho(r)$ via $\rho(r)=A|\Psi_N(\vec{r})|^2$ with the nuclear mass number $A$. 

\begin{figure}[h!]
\centering
  \includegraphics[width=10.0cm,height=7.0cm]{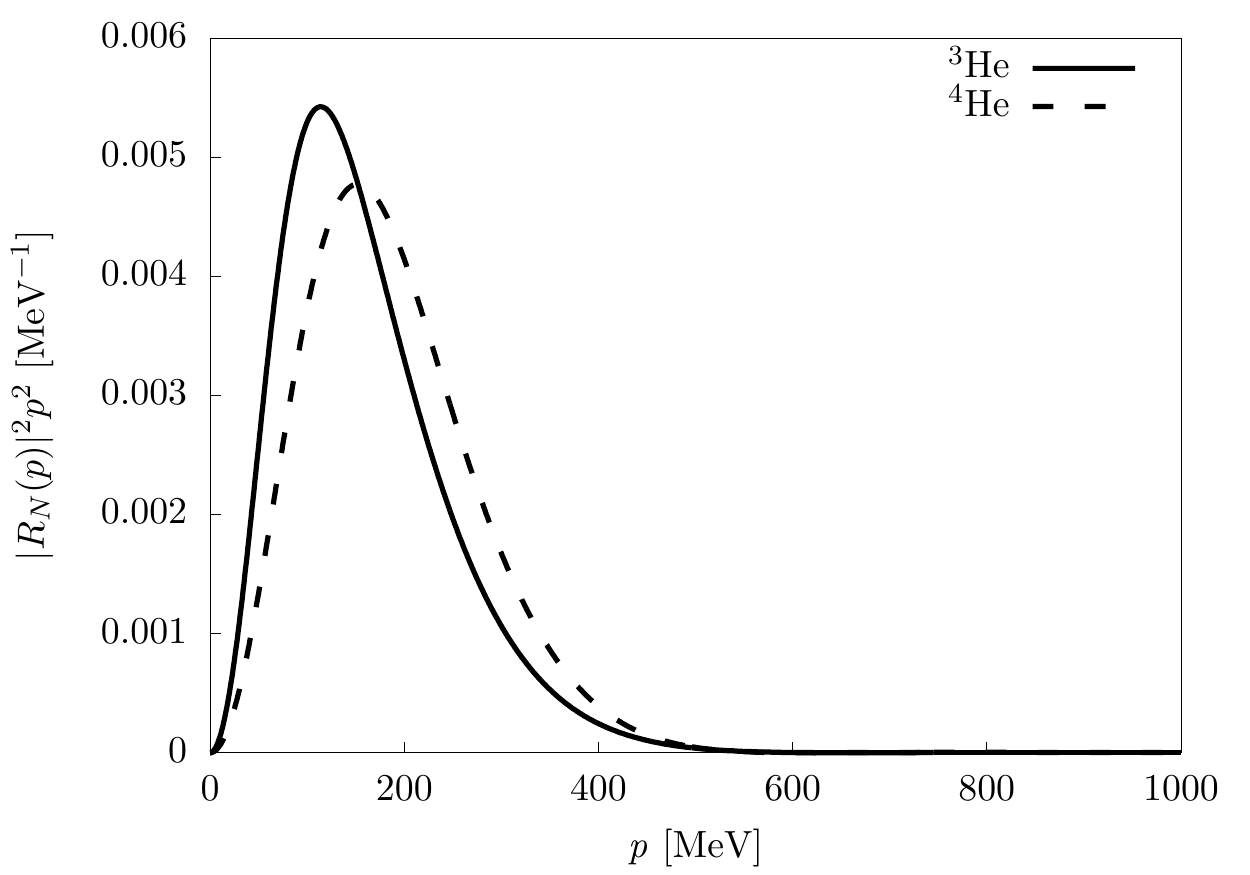}
\vspace{-0.5cm}
 \caption{Fermi momentum distribution of a nucleon in $^3$He (solid line)
 and $^4$He (dotted line) nuclei evaluated by the theoretical nuclear
 density distributions (see details in text).
 \label{fig3}
 }
\end{figure}

\noindent We show in Fig.~\ref{fig3} the calculated momentum
  distribution of the nucleon $|R_N(p)|^2p^2$ based on the $^3$He and $^4$He density distributions~\cite{Hiyama1,Hiyama2,Hiyama3}. 

One can see that in general the momentum distributions for a nucleon inside $^3$He and in $^4$He are broader then the distributions of the relative $^3$He-$\eta$ momentum, though the  distributions for a nucleon inside $^3$He are comparable with the one for $^3$He-$\eta$ for $V_{0}$=-75~MeV.

We can also evaluate the in-medium branching ratios BR$^{*}$ for $\eta
\rightarrow 2\gamma$ and $\eta \rightarrow 3\pi^{0}$ decay channels
below the $\eta$ threshold using the calculated nuclear absorption width
$\Gamma_{\rm abs}$ as,
\begin{equation}
{\rm BR}^{*}_{\eta \rightarrow 2\gamma/\eta \rightarrow
 3\pi^{0}}=\frac{\Gamma_{\eta \rightarrow 2\gamma/\eta \rightarrow
 3\pi^{0}}}{(\Gamma^{\rm tot}_{\eta}+\Gamma_{\rm abs})} \ ,
\label{eq_4}
\end{equation}
%
 where $\Gamma_{\eta \rightarrow 2\gamma/\eta \rightarrow 3\pi^{0}}$ is
 the width of the in-vacuum $\eta$ decay to $2\gamma$ or $3\pi^{0}$, $\Gamma^{\rm tot}_{\eta}$ is the total
 width of $\eta$ meson in vacuum (1.31~keV)~\cite{Pdg2018} and
 $\Gamma_{\rm abs}$ is the nuclear absorption width obtained from Klein-Gordon
 equation.
From the in-vacuum branching ratios BR$_{\eta\rightarrow
 2\gamma/\eta\rightarrow 3\pi^0}$ reported in \cite{Pdg2018},
 $\Gamma_{\eta\rightarrow 2\gamma}$ and $\Gamma_{\eta\rightarrow
 3\pi^0}$ can be calculated as
\begin{eqnarray}
 \Gamma_{\eta\rightarrow 2\gamma}=0.3941\times 1.31\  {\rm keV}\ =0.516\
  {\rm keV}, \\
\Gamma_{\eta\rightarrow 3\pi^0}=0.3268\times 1.31\  {\rm keV}\ =0.428\ 
 {\rm keV}\ .
\end{eqnarray}
%
The estimated branching ratios in medium BR$^*$ are
 listed in the 3rd and 4th column in Table \ref{table_1}. They vary from about 2 $\times$ 10$^{-5}$ to 7 $\times$ 10$^{-4}$ depending on the optical potential parameters. \\ 

The calculated BRs has to be corrected due to the Final State Interaction (FSI) effects of pions with the $^3$He nucleus. We found that the FSI reduces the strength of the signal to half and the effect should be taken into account for the planning and/or analyzing the experiments.

\subsection{Monte Carlo simulation}

The theoretical model described in previous subsection will be applied in the realistic Monte Carlo simulations of the $\eta$-mesic production in the $pd \rightarrow$ $^{3}\hspace{-0.03cm}\mbox{He} 2\gamma$ and $pd \rightarrow$ $^{3}\hspace{-0.03cm}\mbox{He} 6\gamma$ reactions. In the first step, the geometrical acceptance of the WASA detector~\cite{Adam_wasa,Skurzok_PhD} as a function of the excess energy $Q$ near the kinematical threshold for $\eta$ meson production was evaluated. In the simulation, the $^{3}\hspace{-0.03cm}\mbox{He}$ nucleus is assumed to be a spectator and the bound $\eta$ has the energy determined by its mass and binding energy. The decay distribution is assumed to be isotropic in the $\eta$ meson rest frame and the sum of the momenta of the emitted particles are the same as the $\eta$ Fermi momentum in the center of mass frame. In the final state, the $^{3}\hspace{-0.03cm}\mbox{He}$ nucleus has the recoil momentum and energy.  

~The acceptance was determined for the simultaneous registration of $^{3}\hspace{-0.03cm}\mbox{He}$ in the Forward Detector (covering polar angles from 3$^{\circ}$ to 18$^{\circ}$) and $\gamma$ quanta in the Central Detector (covering polar angles from 20$^{\circ}$ to 169$^{\circ}$) and was found to be about 60\% and 40\% for $pd \rightarrow$ $(^{3}\hspace{-0.03cm}\mbox{He}$-$\eta)_{bound}$ $\rightarrow$ $^{3}\hspace{-0.03cm}\mbox{He} 2\gamma$ and $pd \rightarrow$ $(^{3}\hspace{-0.03cm}\mbox{He}$-$\eta)_{bound}$ $\rightarrow$ $^{3}\hspace{-0.03cm}\mbox{He} 6\gamma$ reactions, respectively.

The realistic simulations including the detector responses and all analysis conditions, will be crucial for the interpretation of the experimental data, in particular data collected by WASA-at-COSY Collaboration. The simulation results will be compared with experimental data to choose the most optimal analysis selection conditions as well as to allow the estimation of the overall detection and reconstruction efficiency for the considered processes including all analysis criteria.

\section{Conclusions}

\noindent

In May 2014, WASA-at-COSY Collaboration performed a search for $\eta$-mesic Helium in proton-deuteron collisions. For the first time the hypothetical $^{3}\hspace{-0.03cm}\mbox{He}$-$\eta$ bound state was searched for in non-mesonic $pd \rightarrow$ $^{3}\hspace{-0.03cm}\mbox{He} 2\gamma$ and $pd \rightarrow$ $^{3}\hspace{-0.03cm}\mbox{He} 6\gamma$ decays. For the purpose of the experimental data interpretation, a new theoretical model for the $\eta$-mesic helium was developed, according to which the mesic nucleus decays without $\eta$ meson absorption. The Fermi momentum distribution was determined for a bound $\eta$ meson orbiting around the $^{3}\hspace{-0.03cm}\mbox{He}$ nucleus for different combinations of the $^{3}\hspace{-0.03cm}\mbox{He}$-$\eta$ optical potential parameters. The performed calculations allowed, for the first time, the estimation of the branching ratio for $\eta \rightarrow 2\gamma$ and $\eta \rightarrow 3\pi^{0}$ decay channels in the nuclear medium.

The obtained Fermi momentum distribution will allow the determination of the efficiency for $\eta$-mesic $^{3}\hspace{-0.03cm}\mbox{He}$ production processes and its non-mesonic decays, namely $pd \rightarrow$ $(^{3}\hspace{-0.03cm}\mbox{He}$-$\eta)_{bound}$ $\rightarrow$ $^{3}\hspace{-0.03cm}\mbox{He} 2\gamma$ and $pd \rightarrow$ $(^{3}\hspace{-0.03cm}\mbox{He}$-$\eta)_{bound}$ $\rightarrow$ $^{3}\hspace{-0.03cm}\mbox{He} 6\gamma$.





\section{Acknowledgements}
We acknowledge the support from the Polish National Science Center through grant No. 2016/23/B/ST2/00784. This work was partly supported by JSPS KAKENHI Grant Numbers JP16K05355 (S.H.) and  JP17K05443 (H.N.) in Japan.



\end{document}